\documentclass[manuscript]{aastex63}

\usepackage{txfonts}
\usepackage{latexsym,bm}
\usepackage{multirow}
\newcommand{\qinemail}{qingang@hit.edu.cn}
\newcommand{\hit}{School of Science, Harbin Institute of
Technology, Shenzhen, 518055, China}
\newcommand{\hitqin}{\hit; \qinemail}

\submitjournal{ApJ}

\shortauthors{Qin and Wu}

\turnoffedit

\begin{document}

\title{Magnetic Cloud and Sheath in the Ground-Level Enhancement Event of 2000
July 14. II. Effects on the Forbush Decrease}

\correspondingauthor{G. Qin}
\email{\qinemail}

\author[0000-0002-3437-3716]{G. Qin}
\affiliation{\hitqin}

\author[0000-0002-5776-455X]{S.-S. Wu}
\affiliation{\hitqin}
\begin{abstract}
Forbush decreases (Fds) in galactic cosmic ray intensity are related to
interplanetary coronal mass ejections (ICMEs). The parallel diffusion of particles
is reduced because the magnetic turbulence level in sheath region bounded by ICME's
leading edge and shock is high. Besides, in sheath and magnetic cloud (MC) energetic
particles would feel enhanced magnetic focusing effect caused by the strong
inhomogeneity of the background magnetic field. Therefore, particles would be
partially blocked in sheath-MC structure. Here, we study two-step Fds by considering
the magnetic turbulence and background magnetic field in sheath-MC structure with
diffusion coefficients calculated with theoretical models, to reproduce the Fd
associated with the ground-level enhancement event on 2000 July 14 by solving the
focused transport equation. The sheath and MC are set to spherical caps
\edit1{that are portions of spherical shells}
with enhanced background magnetic field. Besides, the magnetic turbulence levels in
sheath and MC are set to higher and lower than that in ambient solar wind,
respectively. In general, the simulation result conforms to the main characteristics
of the Fd observation, such as the pre-increase precursor, amplitude, \edit1{total}
recovery time, and the two-step decrease of the flux at the arrival of sheath and MC.
It is suggested that sheath played an important role in the amplitude of Fd while MC
contributed to the formation of the second step decrease and prolonged the recovery
time. It is also inferred that both magnetic turbulence and background magnetic
field in sheath-MC structure are important for reproducing the observed two-step Fd.
\end{abstract}


\section{Introduction}
\label{sec:intro}
Forbush decreases (Fds) are short term variations of galactic cosmic ray (GCR)
intensity first observed by \citet{Forbush37} using ionization chambers. Fds can be
classified to two groups, i.e., sporadic Fds and recurrent Fds. A sporadic Fd with
the non-recurrent decrease includes two phases related to the transient
interplanetary coronal mass ejection (ICME), i.e., an impulsive initial phase in
which GCR intensity decreases to its minimum within one day, and a gradual recovery
phase in which GCR intensity recovers to the pre-event level for several days.
Besides, the initial phase of a sporadic Fd shows a two-step decrease sometimes
\citep[e.g.,][]{Cane00, JordanEA11, BhaskarEA16, ShaikhEA17}. The other type of Fds
with the recurrent decrease, caused by high-speed streams from \edit1{coronal} holes,
have gradual decline and recovery phases so that their intensity profiles are more
symmetric \citep[e.g.,][]{Cane00, WawrzynczakEA08, MelkumyanEA19}. In this paper,
the term Fd is used to denote the non-recurrent decrease.

By using neutron monitors, \citet{Simpson54} confirmed that the origin of Fds is in
the interplanetary medium. Before the discovery of ICMEs, some research had already
suggested interplanetary plasma clouds to explain Fds \citep{Forbush37, Morrison56,
CocconiEA58, Piddington58}. Based on the in-situ observations by spacecraft, ICMEs
were found behind the interplanetary shocks in the 1970s, and consequently
\edit1{several research efforts explained}
Fds by considering shocks, turbulent sheath regions, and ICMEs or magnetic clouds
(MCs) that are an important subset of ICMEs \citep{BurlagaEA81, RichardsonEA10}.

\citet{BadruddinEA86} pointed out that MCs may contribute to Fds. Furthermore,
\citet{SandersonEA90} reported that both MCs and turbulent sheath regions are able 
to act as a barrier to cause Fds. In contrast, \citet{ZhangEA88} inferred that MCs
have a little influence on Fds but the main cause is the scattering effect by
turbulent sheath regions. In addition, \citet{LockwoodEA91} argued that MCs don't
have a significant effect on GCRs while the reduced diffusion in turbulent sheath
region has a large contribution to Fds. By using the isotropic intensity of GCRs
provided by the \textit{IMP 8}, \citet{Cane93} confirmed that MCs can cause the
decrease of GCR intensity. Besides, \citet{CaneEA97} and \citet{RichardsonEA99} 
found that even small ICMEs can produce a signal of GCR decreases at 1 au.
\citet{RichardsonEA11} carried out a statistical study of over 300 ICMEs and
exhibited that 80\% of the ICMEs were associated with Fds, and they also concluded
that the maximum depths of Fds caused by MCs are much deeper than that by the 
non-MC ICMEs. \citet{JordanEA11} found that the small-scale magnetic structures in
sheath regions can modulate GCR intensities significantly. What's more,
\citet{YuEA10} and \citet{ArunbabuEA15} showed that the enhanced magnetic turbulence
level and background magnetic field in sheath region contribute to the formation 
of Fds. Note that the total magnetic field is the superposition of the background
magnetic field and the turbulent magnetic field.

\edit1{Observational studies were also devoted to exploring the effect of the
polarity states of the heliospheric magnetic field on Fds. The polarity $A$
determines the direction of drift velocity, which is usually used to explain the
polarity dependence of Fds in the literature. Most of these research statistically
compared Fd characteristics during different polarity conditions.
\citet{LockwoodEA86} observed no significant difference in the characteristic
recovery time, defined as the time for the decrease to decay up to $e^{-1}$ times
its amplitude, with the reversal of the heliospheric magnetic field. In contrast,
\citet{RanaEA96} and \citet{SinghABadruddin06} found that the characteristic
recovery time is longer during $A<0$ epoch than that during $A>0$ epoch.
\citet{RanaEA96} also observed no significant difference in the percentage of recovery
up to the 10th day during the two polarity states, which indicates that the total
recovery time may be less influenced by particle drift than the characteristic recovery
time. Besides, \citet{SinghABadruddin06} showed that the amplitude of Fds is not
significantly different during $A>0$ and $A<0$ polarity conditions.
\citet{MulderAMoraal86} found that there is a small drift effect on Fd profiles,
especially on small ones, i.e., the reset time, defined as the time at which Fds
have reset to 75\%, 50\%, 37\% ($e^{-1}$), and 25\% of their amplitude, is longer
for the situation $A<0$ than that for the situation $A>0$. They considered that the
background heliospheric magnetic field configuration responsible for the
drift is essentially wiped out because of the large blast wave so that the reset
time is not dependent on the polarity for large Fds.}

\edit1{Increasingly complex numerical simulations have been developed}
for several decades. By solving a 1D diffusion-convection transport equation,
\citet{Nishida82} studied the precursory increase during an Fd event. Furthermore,
\citet{KadokuraEA86} developed a 2D GCR transport model to study Fds by using a
diffusion barrier. In addition, \citet{ThomasEA84} studied the dependence of the
maximum intensity decrease and recovery time of Fds on diffusion coefficient,
particle rigidity, and flare geometry with the Monte-Carlo simulations, to report
that the geometry of flare compression can significantly affect the maximum intensity
decrease and recovery time of Fds. By considering the propagating diffusion barrier
as the main cause of Fds and solving a 2D transport equation where adiabatic cooling
and particle drift are included, \citet{leRouxEA91} showed the effect, which could
be weakened if the tilt angle of heliospheric current sheet (HCS) gets increase,
that the recovery time of Fds is \edit1{longer} when the polarity of the solar
magnetic field is negative than that when the polarity is positive.
\edit1{Besides, their simulation indicated that the Fd amplitudes are almost
equal during different polarity conditions in contrast to the results of
\citet{KadokuraEA86}.}
Recently, \citet{LuoEA17, LuoEA18} carried out a 3D simulation of proton and
electron Fds based on a stochastic differential equation approach adopting the
scenario that Fds are mainly caused by propagating diffusion barriers too.
\edit1{Their results inferred that the characteristic recovery time of proton Fd is
longer/shorter than that of electron one when the polarity of the heliospheric
magnetic field is negative/positive, while there is little charge-sign dependent
effect on the amplitude of Fds.}

From the above discussion it is suggested that there are contradictions between
observations and simulations for Fds. On the one hand, observational studies showed
that the sheath region and MC are able to cause Fds which have two-step decrease
sometimes. On the other hand, the simulation studies in the literature have been
concentrated on producing Fds by using diffusion barrier, in which the diffusion is
reduced artificially, with only one-step decrease generated. Furthermore, the
ground-level enhancement (GLE) events of solar energetic particles (SEPs), usually
accompanying with large and fast ICMEs \citep[e.g.,][]{GopalswamyEA12} which drive
strong ICME shocks, are of great interest to researchers
\citep[e.g.,][]{GopalswamyEA12, MewaldtEA12, WuAQin18, WuAQin20, FirozEA19}. In the
GLE59 \edit1{that} occurred on 14 July 2000, there was a very fast and strong MC
\citep{Lepping01}, for which event in \citet[][hereafter, WAQ-I]{WuAQin20} we
studied the effect of sheath and MC on SEPs accelerated by the ICME shock through
numerically solving the focused transport equation, with the background magnetic
field and magnetic turbulence levels in sheath and MC quite different from that in
ambient solar wind, and diffusion coefficients calculated with diffusion theories. 
It is shown that a two-step Fd were observed when the ICME-shock structure
associated with GLE59 arrived at the Earth. In this work, as a continuation of 
WAQ-I, with the similar model for the sheath-MC structure and diffusion 
coefficients, we numerically study the Fd to reproduce the two-step decrease. The
simulation results of WAQ-I showed that the sheath-MC structure reduced the proton
intensities for about 2 days after shock passing through the Earth. It was further
found that the sheath contributed most of the decrease while the MC facilitated the
formation of the second step decrease. The simulation also inferred that the
\edit1{combination} of background magnetic field and magnetic turbulence in sheath-MC
structure can produce a stronger effect of reducing SEP intensities. The
observations of the Fd associated with GLE59 are presented in Section~\ref{sec:obs}.
The simulation model is elaborated in Section~\ref{sec:model}. We show the simulation
results in Section~\ref{sec:result}. And conclusions and discussion are presented in
Section~\ref{sec:discussion}.

\section{OBSERVATIONS}
\label{sec:obs}
\begin{figure}
\epsscale{1} \plotone{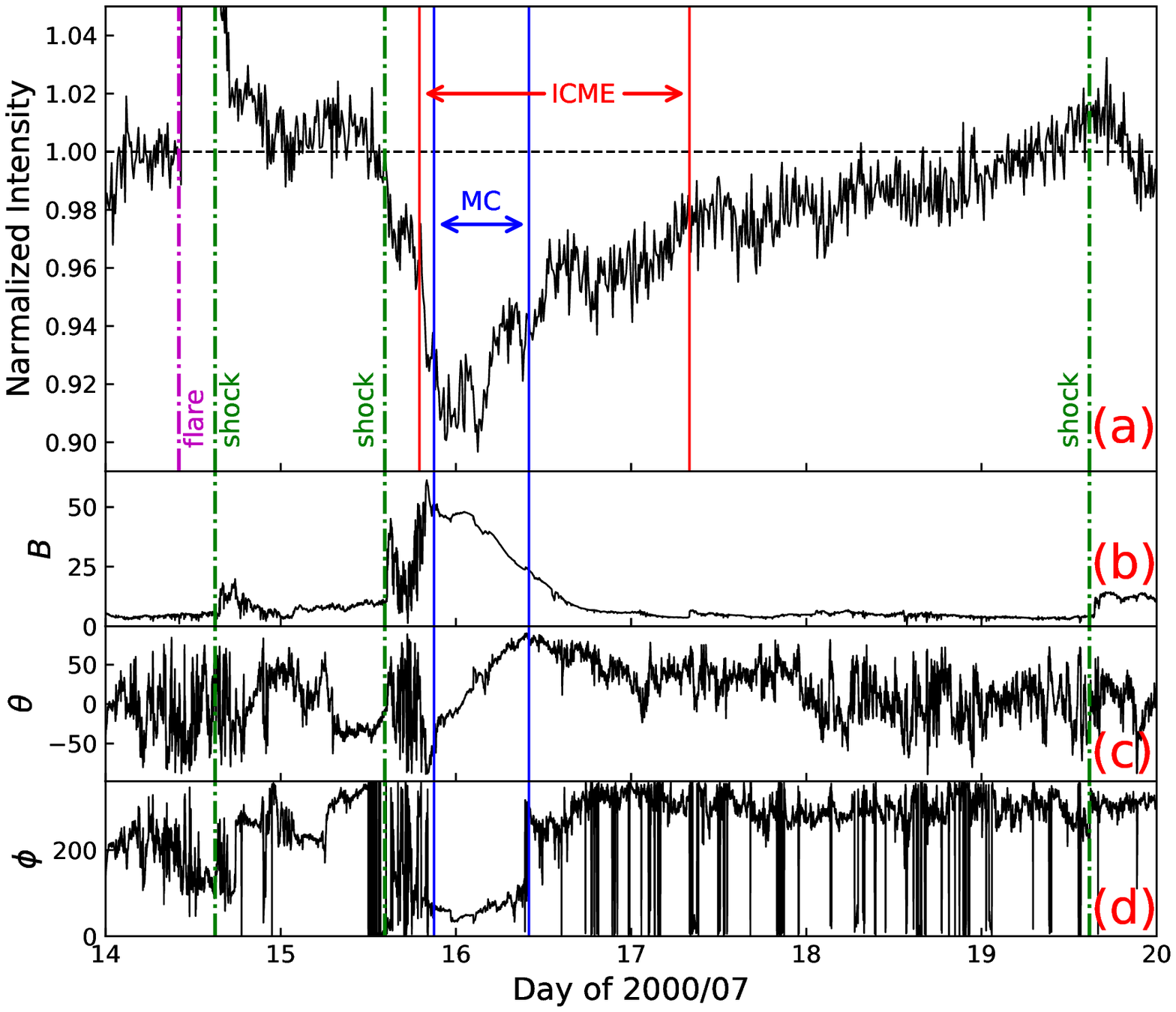}
\caption{Observations for the Fd associated with GLE59. (a) The normalized intensity
of GCR measured by the Oulu NM is plotted with the black solid curve, and the
pre-event level is presented with the black horizontal dashed line. The pink and
green vertical dashed lines denote flare onset and the passages of ICME shock,
respectively. The boundaries of ICME and MC are presented with the red and blue
vertical solid lines, respectively. (b)$-$(d) are the intensity, polar angle, and
azimuthal angle of IMF provided by \textit{Wind} spacecraft in GSE angular
coordinates, respectively.}
\label{fig:observation}
\end{figure}

Figure~\ref{fig:observation} shows the observations of the Fd associated with GLE59.
Panel (a) is the normalized count rate from the Oulu neutron monitor (NM). Panels
(b), (c) and (d) present the intensity, polar angle, and azimuthal angle of
interplanetary magnetic field (IMF) in GSE angular coordinates from the \textit{Wind}
spacecraft, respectively. In Panel (a), there was an X5.7 class flare that began at
10:10 UT on 14 July 2000 indicated by the pink vertical dashed line and located at
N22W07, there were also three interplanetary shocks that arrived at the Earth denoted
by the green vertical dashed lines and the second shock corresponded to the solar
eruption. An ICME behind the second shock was observed at the Earth with start and end
times 19:00 UT on 15 July and 8:00 UT on 17 July indicated by the two red vertical
lines, respectively. The ejecta is believed to be a MC \citep{RichardsonEA10} with
boundaries at 21:00 UT 15 July and 10:00 UT 16 July denoted by the two blue vertical
lines. After the flare onset, the cosmic ray intensity had an impulsive increase that
is called GLE59, and then the cosmic ray intensity dropped rapidly to the pre-event
intensity \edit1{(normalized to a value of 1)}
for about half a day. When the flare corresponding shock arrived at the Earth, the
GCR intensity dropped rapidly to about 0.97. In addition, the GCR intensity decreased
to nearly 0.9 when the MC's leading edge arrived at the Earth. After the two-step
decrease, the GCR intensity recovered gradually for several days.
Figure~\ref{fig:observation}(b) exhibits that the sheath region between the ICME
shock and ICME's leading edge had an impulsive magnetic field enhancement right
behind the ICME shock. Besides, the background magnetic field enhanced again when
the ICME arrived at the Earth and the enhancement lasted until the MC left the Earth.
What's more, the magnetic field observed at the Earth also indicates that the
magnetic turbulence levels in the sheath and MC were higher and lower than that in
ambient solar wind, respectively.

From Figure~\ref{fig:observation} it is thus shown that the occurrence of two-step
decrease of GCR intensity coincided with the arrivals of sheath and MC, so that we
assume the two-step Fd associated with GLE59 was caused by the sheath-MC structure. 

\section{GCR Transport MODEL}
\label{sec:model}

\subsection{Transport Equation}
The Parker transport equation \citep{Parker65} is widely used to study the modulation
of GCRs
\begin{equation}
	\frac{{\partial f}}{{\partial t}} =
	-\left( \bm{V}^{\text{sw}} + \left< \bm{v}_{\text{d}} \right> \right) \cdot \nabla f +
	\nabla \cdot \left( \bm{K}_{\text{s}} \cdot \nabla f \right) 
	+\frac{1}{3}\left( \nabla \cdot \bm{V}^{\text{sw}} \right) 
	\frac{\partial f}{\partial \ln{p}},
	\label{eq:Parker}
\end{equation}
where $f(\bm{x},p,t)$ is the omnidirectional particle distribution function, with
$\bm{x}$ the particle position in a \edit1{3D} non-rotating heliographic coordinates,
$p$ the particle momentum, and $t$ the time. The first term on the right hand side
represents the solar wind flowing and particle drift in in-homogeneous IMF, with
$\bm{V^{\text{sw}}}=V^{\text{sw}}\bm{e_r}$ the solar wind velocity in radial
direction, and $\left< \bm{v}_{\text{d}} \right>$ the pitch-angle averaged drift
velocity. The second term on the right hand side refers to the diffusion effects,
with $\bm{K}_{\text{s}}$ the symmetric part of diffusion tensor. The last term on
the right hand side is the adiabatic energy losses.

WAQ-I suggested that the rapid changed magnetic fields in sheath and MC act as
magnetic mirrors due to the magnetic focusing effect, blocking the passage of SEPs,
which depends on particle's pitch-angle cosine $\mu$. However, pitch-angle cosine has
been eliminated in Equation~(\ref{eq:Parker}) assuming GCRs to be isotropic. In this
work, we therefore choose the focused transport equation rather than the Parker
equation for modeling the transport of GCRs in the presence of sheath and MC.
The focused transport equation is given by
\citep{Skilling71, Schlicheiser02, QinEA06, ZhangEA09, ZhangEA19}
\begin{eqnarray}
\frac{{\partial f}}{{\partial t}} + 
\left( v \mu \hat{\bm{b}} + \bm{V}^{\text{sw}} +\bm{v}_{\text{d}} \right) \cdot \nabla f -
  \nabla \cdot \left( \bm{\kappa_\bot} \cdot \nabla f \right) - 
  \frac{\partial}{{\partial \mu}} 
  \left( D_{\mu \mu} \frac{{\partial f}}{{\partial \mu}} \right) \nonumber \\
- p \left[ {\frac{{1 - \mu^2}}{2} \left( {\nabla \cdot \bm{V}^{\text{sw}} -
  \hat{\bm{b}} \hat{\bm{b}} : \nabla \bm{V}^{\text{sw}}} \right) +
  \mu^2 \hat{\bm{b}} \hat{\bm{b}} : \nabla \bm{V}^{\text{sw}}} \right]
  \frac{{\partial f}}{{\partial p}} \nonumber \\
+ \frac{{1 - \mu^2}}{2} \left[ {-\frac{v}{L} + \mu \left( 
  \nabla \cdot \bm{V}^{\text{sw}} - 
  3 \hat{\bm{b}} \hat{\bm{b}} : \nabla \bm{V}^{\text{sw}} \right) } \right]
  \frac{{\partial f}}{{\partial \mu}}=0,
\label{eq:Fokker}
\end{eqnarray}
where $f(\bm{x},\mu,p,t)$ is the gyrophase-averaged distribution function, $v$ is
the speed of particles, $\bm{\kappa_\bot}$ and $D_{\mu\mu}$ are the perpendicular
and pitch-angle diffusion coefficients, respectively,
$L = \left( \hat{\bm{b}} \cdot \nabla \ln{B_0} \right)^{-1}$ is the magnetic focusing
length, $\hat{\bm{b}}$ is the unit vector along the local magnetic field, and $B_0$
the strength of local magnetic field.

The numerical solution of Equation~(\ref{eq:Fokker}) needs more computing resources
than that of Equation~(\ref{eq:Parker}) because there is one more independent
variable, $\mu$. \edit1{On} the other hand, Fd is a short term process caused mainly
by local structures to last several days, during which we focus on the relative
variation of GCRs, so that the outer boundary is set to a symmetric spherical
boundary at 10 au \citep[e.g.,][]{Zhang99} instead of 85 au
\citep[e.g.,][]{QinAShen17} or beyond \citep[e.g.,][]{PotgieterEA14}, for saving of
the computing resources, where the source of GCRs can be written as
\citep{Zhang99, ShenAQin18, ShenEA19}
\begin{equation}
f_{\text{s}} = \frac{f_0 p_0^{2.6}}{ p\left(m_0^2 c^2 + p^2 \right)^{1.8}},
 \label{eq:fb}
\end{equation}
where $f_0$ is a constant, $m_0$ is the mass of protons, and $p_0=1$GeV/$c$.
In Figure~\ref{fig:MC_sheath}(a) we plot the energy spectrum of GCR source
calculated from the model of Equation~(\ref{eq:fb}) in arbitrary units at 10 au
with the black line.

\edit1{As discussed in Section~\ref{sec:intro}, in observations, the characteristic
recovery time is, on average, longer during $A<0$ epoch than that during $A>0$ epoch.
However, there is no clear polarity dependent effect on the reset time for large Fds.
Besides, the total recovery time may be less influenced by particle drift than the
characteristic recovery time. In simulations, the polarity dependent effect on the
characteristic recovery time could be weakened if the tilt angle of HCS gets
increase. Furthermore, the particle drift may have no significant effect on the
amplitude of Fds because there is little polarity dependent effect on the amplitude
of Fds both in observations and simulations. In this work, both the Fd amplitude and
the tilt angle of HCS in the Fd event that occurred following the GLE59 on 2000 July
14 are large. Therefore, we assume that the diffusion and magnetic mirror effects in
the sheath-MC structure are more important than the drift effect in forming the Fd
associated with GLE59, so that we neglect the drift term $\bm{v}_{\text{d}}$ in our
simulation for focusing on the formation of the two-step decrease, amplitude, and
total recovery time of the Fd, although the particle drift may affect the
characteristic recovery time slightly in this event.}

In order to compare the GCR count rate provided by NMs with flux from simulation
model, the effective energy of NMs is used \citep{AlankoEA03, ZhaoEA14}
\begin{equation}
E_{\text{eff}} = E_1 + \frac{ E_2 \left( \frac{P_{\text{c}}}{P_1} \right)^{1.25} }
  { 1 + 10 \exp{ \left( -0.45\frac{P_{\text{c}}}{P_1} \right) }},
\label{eq:E_eff}
\end{equation}
where $P_{\text{c}}$ is the local geomagnetic cutoff rigidity of NMs, and $E_1$,
$E_2$, and $P_1$ are constants with $E_1=6.4$ GeV, $E_2=1.45$ GeV, and $P_1=1$ GV. 
The count rate, $N$, is proportional to the integral GCR flux above the effective
energy, i.e.,
\begin{equation}
N \propto \int_{E_{\text{eff}}}^{+\infty} j(E)dE,
\end{equation}
where $j(E)=p^2 f$ is the differential flux and $f$ is obtained from
Equation~(\ref{eq:Fokker}). Following the previous studies \citep{QinEA06, ZhangEA09},
we use a time-backward Markov stochastic process method \citep{Zhang99} to
numerically solve Equation~(\ref{eq:Fokker})
\edit1{in a 3D heliocentric coordinate system}
to obtain the anisotropic distribution function $f(\bm{x},\mu,p,t)$. In addition,
to average $f(\bm{x},\mu,p,t)$ over $\mu$ we can finally get the isotropic
distribution function $f(\bm{x},p,t)$. Here, the position $\bm{x}$ is set to Earth.

\subsection{IMF, MC, and sheath}
In this work, we adopt the similar model of the IMF, MC, and sheath as in WAQ-I.
Parker field is adopted as the background solar wind magnetic field
\begin{equation}
\bm{B}_{\text{P}} = A B_{\text{P}0} \left( \frac{r_{\text{au}}}{r} \right)^2 
 \left(\bm{e_r} - \frac{\omega r \sin{\theta}}{V^{\text{sw}}} \bm{e_{\phi}}\right),
 \label{eq:BP}
\end{equation}
where $B_{\text{P}0}$ is a constant and equal to the radial strength of the
background magnetic field at 1 au without a local structure, $r_{\text{au}}$ equals
to 1 au, $\omega$ is the angular speed of solar rotation, and $r$, $\theta$, and
$\phi$ are the solar distance, \edit1{colatitude}, and longitude of any point,
respectively. Note that, the HCS latitudinal extent is not included in
Equation~(\ref{eq:BP}) because we neglect the drift effect.

\begin{figure}
\epsscale{1} \plotone{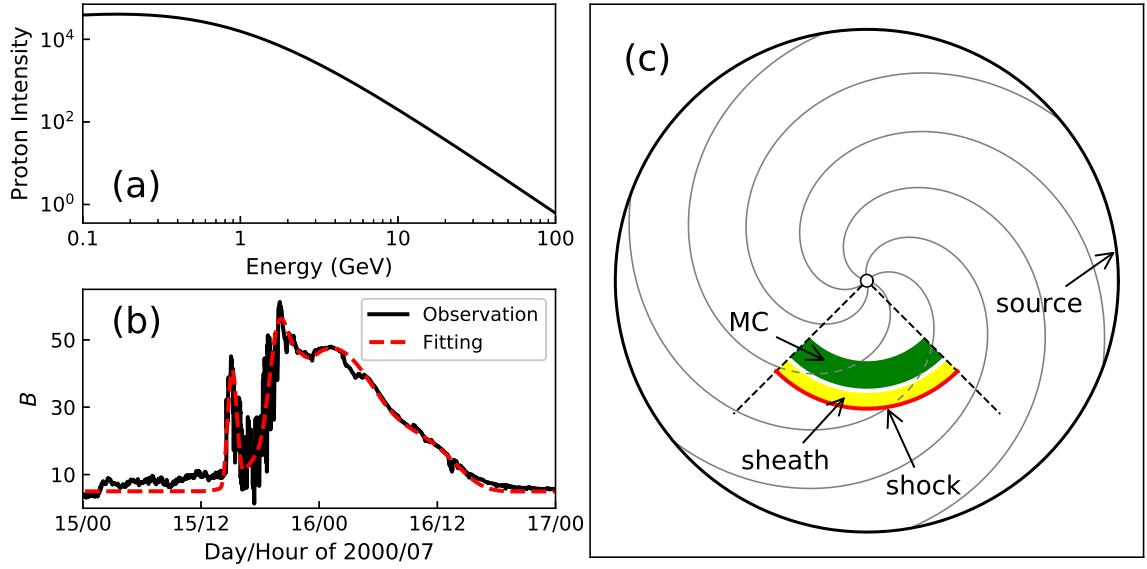}
\caption{(a) The energy spectrum of GCR source in arbitrary units at 10 au is
plotted with the black line. (b) Fitting (red dashed line) of observed background
magnetic field (black solid line) in sheath-MC structure at 1 au with
Equation~(\ref{eq:Bejecta}). (b) A sectional view of the IMF (gray spiral lines),
shock (red arc), sheath (yellow cap), MC (green cap), and GCR source (black circle)
through the ecliptic plane.}
\label{fig:MC_sheath}
\end{figure}

In this work, the ICME shock, assumed not affecting the transport of GCRs, is used
as a reference to determine some parameters of the sheath and MC. The shock is
\edit1{modeled as} a spherical cap \edit1{that is a portion of a spherical shell}
with a uniform speed that is obtained by dividing the Sun-Earth distance by the
shock transit time, and the direction of nose in the flare location. The MC and
sheath are set to thick spherical caps that are parallel to the ICME shock with
thicknesses to be fixed at the observed ones at 1 au, and speeds and angular widths 
the same as that of the shock. Due to the fact that the magnetic field in sheath-MC
structure is higher than that in ambient solar wind as shown in
Figure~\ref{fig:observation}(b), the background magnetic field in sheath-MC 
structure, $\bm{B_{\text{ejecta}}}$, is simply set to an ejecta model following
WAQ-I, i.e., Parker field plus a magnetic field enhancement in radial direction
\begin{equation}
\bm{B_{\text{ejecta}}} = \bm{B_{\text{P}}} + A \Delta B_r \bm{e_r},
 \label{eq:Bejecta}
\end{equation}
where $\Delta B_r$ can be expressed by the sum of a set of delta-like functions
\begin{eqnarray}
\Delta B_r &=& \sum_{i=1}^k \Delta B_{r}^i, \label{eq:deltaBr}\\
\Delta B_r^i &=& B_{r0}^i \left(\frac{r_{\text{au}}}{v_{\text{s}}t}\right)^2 
 \delta_n \left( \frac{v_{\text{s}} t + \delta r_i -r}{w_i} \right), \label{eq:deltaB}\\
\delta_n(x) &=& \left\{
\begin{array}{ll}
\left(1-x^2\right)^n & \qquad \text{for } x \in \left[-1,1\right],\\
0 & \qquad \text{for others},
\end{array} \right. \label{eq:deltan}
\end{eqnarray}
here $B_{r0}^i$, $\delta r_i$, $w_i$ ($i=1, 2, 3, ...,k$), and $n$ are constants  
obtained by fitting the observed background magnetic field with 
Equation~(\ref{eq:Bejecta}), and $v_{\text{s}}$ is the shock speed. The fitting 
result of the background magnetic field is presented in Figure~\ref{fig:MC_sheath}(b)
where the black and red curves are the observed and fitted background magnetic field,
respectively. The detailed fitting coefficients can be found in the Figure 2(a) and
Table 1 of WAQ-I. Note that, the polar and azimuthal angles of the fitted magnetic
field are inconsistent with those of the observed one, which is discussed in WAQ-I.
Therefore, the background magnetic field $\bm{B}_0$ is written as
\begin{equation}
    \bm{B}_0= \left\{
\begin{array}{ll}
\bm{B}_{\text{P}} & \qquad \text{in solar wind},\\
\bm{B_{\text{ejecta}}}  & \qquad \text{in sheath-MC}.
\end{array} \right.
\end{equation}

Figure~\ref{fig:MC_sheath}(c) presents the sectional view of the IMF, shock, MC,
sheath, and GCR source through the ecliptic plane with the gray spiral curves, red
arc, thick green cap, thick yellow cap, and black circle, respectively. The dashed
spiral curves are used to denote that the background magnetic field in sheath-MC
structure is not Parker field. It is noted that Figure~\ref{fig:MC_sheath}(c) shows
similar models of the IMF, shock, MC, and sheath as Figure 2(b) of WAQ-I, except the
GCR source boundary which is added for Fd study here.

The magnetic turbulence is based on a two-component 2D+slab model with turbulence 
level given by
\begin{equation}
\sigma \equiv \frac{\delta b}{B_0} = 
 \frac{\sqrt{\delta b_{\text{slab}}^2 + \delta b_{\text{2D}}^2}}{B_0},
\end{equation}
where $\delta b_{\text{slab}}$ and $\delta b_{\text{2D}}$ are the slab and 2D
components of magnetic turbulence, respectively. The ratio of 2D energy to slab 
energy is found to be 80\%:20\% \citep{MatthaeusEA90, BieberEA94}, so that the
turbulence levels of slab and 2D components in solar wind, sheath, and MC can be
written as
\begin{eqnarray}
\left( \frac{\delta b_{\text{slab}}}{B_0} \right)_i &=& 
  \frac{\sqrt{5}}{5}\sigma_i  \qquad (i=\text{P, S, M}),\\
\left( \frac{\delta b_{\text{2D}}}{B_0} \right)_i &=& 
  \frac{2\sqrt{5}}{5}\sigma_i  \qquad (i=\text{P, S, M}),
\end{eqnarray}
where $i=\text{P, S, M}$ is used to denote solar wind, sheath, and MC. The values of
$\sigma_{\text{S}}$ and $\sigma_{\text{M}}$ should be set to higher and lower than
that of $\sigma_{\text{P}}$ due to the fact that the magnetic turbulence level in
sheath and MC are greater and less than that in solar wind, respectively. We can
also collectively refer $\sigma_{\text{S}}$ and $\sigma_{\text{M}}$ as the ejecta
model $\sigma_{\text{ejecta}}$.

\subsection{Diffusion Coefficients}
As in WAQ-I, the diffusion coefficients are obtained with the models in the
following. The pitch-angle diffusion coefficient $D_{\mu \mu}$ in
Equation~(\ref{eq:Fokker}) has the expression given by
\citep{BeeckEA86, TeufelEA03}
\begin{equation}
{D_{\mu \mu}}(\mu) = {\left( {\frac{{\delta {b_{\text{slab}}}}}{{{B_0}}}} \right)^2}
 \frac{{\pi (s-1)}}{{4s}} \frac{v}{l_{\text{slab}}} \left(\frac{R_L}{l_{\text{slab}}}\right)^{s-2}
 \left({\mu^{s - 1}} + h \right) \left(1 - {\mu^2} \right),
 \label{eq:Dmumu}
\end{equation}
where $s=5/3$ is the Kolmogorov spectral index of the IMF turbulence in inertial 
range, $l_{\text{slab}}$ is the correlation length of the slab component of
turbulence, $R_L = pc / \left( |q| B_0 \right)$ is the Larmor radius, $\mu$ is
pitch-angle cosine, and $h=0.01$ is introduced to model the non-linear effect of
pitch-angle diffusion at $\mu=0$. 

The other diffusion coefficient, perpendicular diffusion coefficient 
$\bm{\kappa_{\bot}}$, in Equation~(\ref{eq:Fokker}) is from non-linear guiding
center theory \citep{MatthaeusEA03} with analytical approximations
\citep{ShalchiEA04, ShalchiEA10}
\begin{equation}
\bm{\kappa_{\bot}} = 
\frac{v}{3} {\left[ {{{\left( \frac{{\delta {b_{\text{2D}}}}}{{{B_0}}} \right)}^2}
\sqrt {3\pi } \frac{{s - 1}}{{2s}}
\frac{{\Gamma \left( \frac{s}{2} + 1 \right)}}
{{\Gamma \left( \frac{s}{2} + \frac{1}{2} \right)}}
{l_{\text{2D}}}} \right]^{2/3}} {\lambda_\parallel}^{1/3}
\left( \bm{I} - \bm{\hat{b}} \bm{\hat{b}} \right),
\end{equation}
where $l_{\text{2D}}$ is the correlation length of the 2D component of turbulence,
$\bm{I}$ is a unit tensor, and $\lambda_{||}$ is the parallel mean free path
\citep{Jokipii66, HasselmannEA68, Earl74}
\begin{equation}
{\lambda_\parallel} = \frac{{3v}}{8} \int_{-1}^{+1} 
  {\frac{\left( 1 - {\mu^2} \right)^2}{D_{\mu \mu}} d \mu}.
\end{equation}
We can also collectively refer $l_{\text{slab}}$ and
$l_{\text{2D}}$ as $l_{\text{turb}}$.

\section{SIMULATIONS AND COMPARISONS WITH OBSERVATIONS}
\label{sec:result}

\subsection{Parameter Settings}
\label{sec:para}
The main parameters of the simulations are listed in Table~\ref{tab:para}. According 
to the observed times of the flare onset and shock passage at 1 au, the shock speed
is set to 1406 km/s. The half angular width of the shock, $\Omega_{\text{s}}$, is set
to 45$^\circ$. The solar wind speed is set to 450 km/s, and the radial strength of
IMF at 1 au, $B_{\text{P}0}$, is set to 3.62 nT, making the total strength of IMF, 
$B_{\text{P}}|_{1\text{au}}$, equal to 5 nT at 1 au. According to the observed start
and end times of MC and sheath, the half thicknesses of sheath and MC are set to 0.08
au and 0.22 au, respectively, and the distance from the ICME shock to the center of
MC, $d_{\text{M}}$, is set to 0.45 au. The parameters about magnetic turbulence are
listed in Table~\ref{tab:para_turbulence}. We set $l_{\text{slab}}=0.025$ au, so that
$l_{\text{2D}}$ equals to $l_{\text{slab}}/2.6=0.0096$ au according to the
multi-spacecraft measurements \citep{WeygandEA09, WeygandEA11}. The magnetic
turbulence levels in solar wind, sheath, and MC are set to 0.3, 1.6, and 0.1,
respectively. 
\edit1{Note that, since it is not easy to measure turbulence levels in transient
region of solar wind, sheath, and MC accurately, we set the values of turbulence
levels with the assumption to produce good results.}
The other parameters such as the angular speed of solar rotation is set to
$\omega=2\pi/25.4$ rad/day, the inner and outer boundaries of the simulations are 
set to $R_{\text{in}}=0.05$ au and $R_{\text{out}}=10$ au, respectively, and the
effective energy of the Oulu NM obtained from Equation~(\ref{eq:E_eff}) equals to
6.54 GeV. Note that all the parameters in Tables \ref{tab:para} and
\ref{tab:para_turbulence} except the last two parameters in Table \ref{tab:para},
which are specific for GCRs, are the same as that in Tables 1 and 2 of WAQ-I. On the
other hand, here we do not include the parameters for the source of SEPs moving with
the shock that are in Table 1 of WAQ-I.

\subsection{Results}
\begin{figure}
\epsscale{1} \plotone{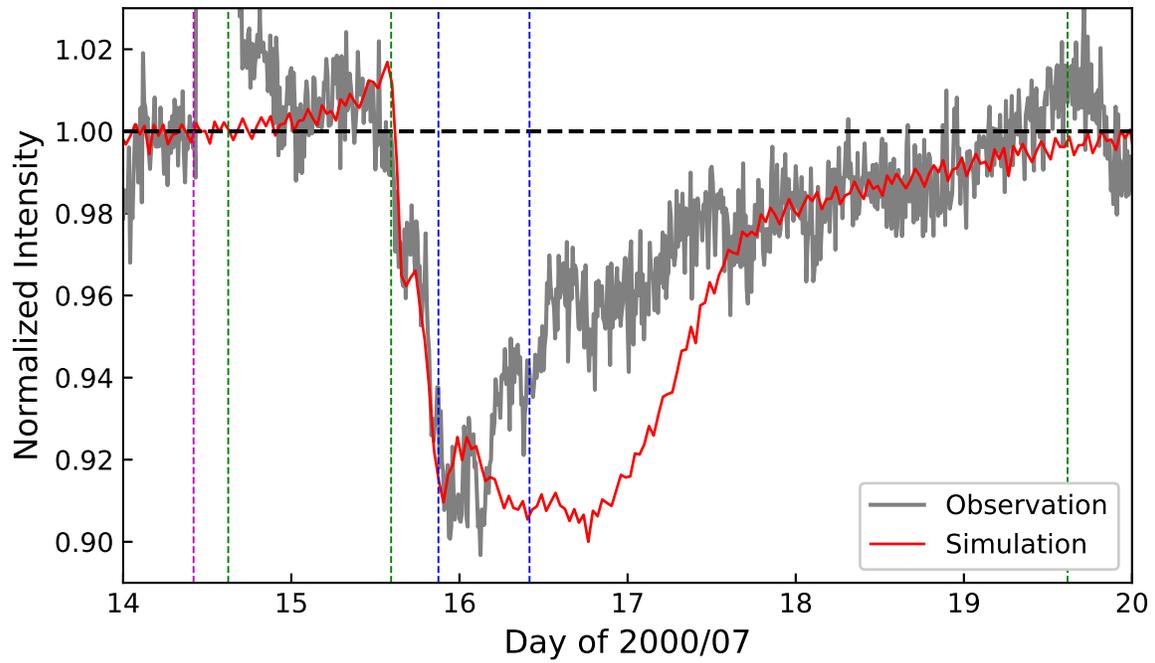}
\caption{Observation and simulation of the Fd are plotted with gray and red curves,
respectively. The black horizontal dashed line denotes the pre-event level of GCR
intensity. The pink, green, and blue vertical dashed lines represent flare onset,
shock passages, and MC boundaries, respectively.}
\label{fig:simulation}
\end{figure}

The data from observation and simulation result are presented with gray and red
curves, respectively, in Figure~\ref{fig:simulation} with the pre-event level of GCR
intensity indicated by the black horizontal dashed line. In addition, the pink,
green, and blue vertical dashed lines denote the flare onset, shock passages of the
Earth, and MC boundaries, respectively. It is shown that the simulation result fits
the observed decrease phase well to reproduce the two-step decrease. The pre-increase
precursor of Fd, which can be found in some Fds, is also reproduced in the simulation
result. From the observations and simulations we can see that the pre-increase
started about 15 hours before the
\edit1{arrival of the shock, and the time scale is similar to the statistical
result from observation, which is 10$-$14 hours in average}
\citep{LingriEA19}. Though the observation of the pre-event GCR intensity has been
contaminated by GLE59, the pre-increase in observation can be recognized roughly and
is consistent with the simulation. The recovery phase of simulation, however,
deviates with that of the observation during the time period between July 16 and
July 18, after which the simulation is consistent with the observation again. At the
end of the Fd, the observed GCR intensity was influenced by another ICME that was
behind the third shock, so that the \edit1{total} recovery time is also affected.
If the observed GCR intensity is not contaminated, it can be suggested that the
\edit1{total} recovery time may equal to about 4 days, which is in accordance with
that of simulation.

\begin{figure}
\epsscale{1} \plotone{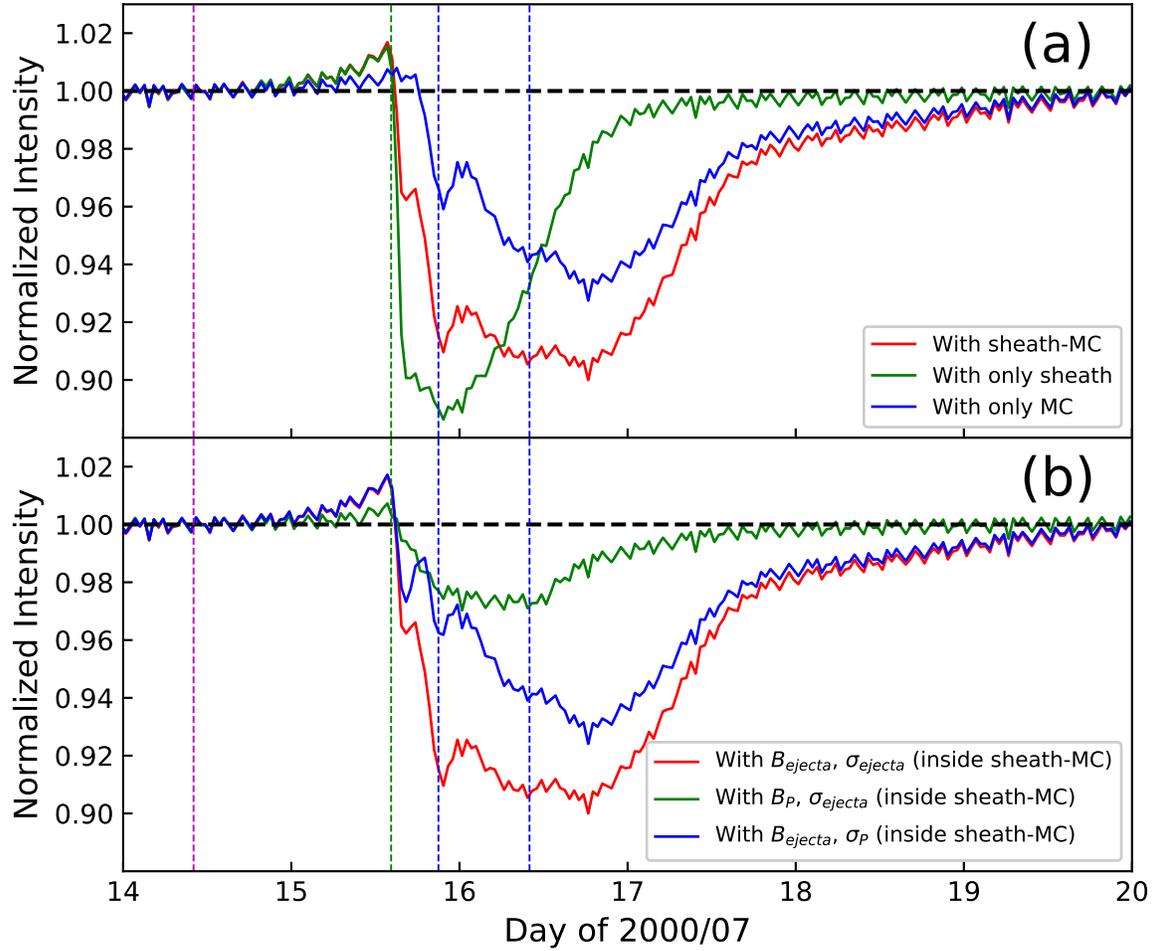}
\caption{(a) Simulation results with only sheath and only MC are presented by the 
green and blue curves, respectively. (b) Simulation result with sheath-MC but only
magnetic turbulence level in sheath-MC structure is different from that of ambient
solar wind is plotted with green curve, while simulation result with sheath-MC but 
only background magnetic field is different from that of ambient solar wind is
presented by blue curve. Other lines are the same as that in
Figure~\ref{fig:simulation}.}
\label{fig:simulation_compare}
\end{figure}

To evaluate the influence of sheath and MC, we further run simulations with only
sheath or MC, with the results presented by the green and blue curves in
Figure~\ref{fig:simulation_compare}(a), respectively. The other curves are the same
as that in Figure~\ref{fig:simulation}. It is shown that the simulation with only 
sheath, i.e., the green curve, has a sharp decrease right after the shock arrival, 
with an amplitude slightly larger than that of the simulation with sheath-MC, i.e.,
the red curve. The green curve recovers more rapidly with the \edit1{total} recovery
time about the half of that of the red curve. The simulation with only MC, i.e., the
blue curve, has a fairly rapid decrease when MC's leading edge arrives at the Earth
followed by a slow decrease lasting for about 1 day, with the smallest amplitude.
Afterwards, the blue curve recovers gradually with the recovery phase in line with
the red curve in the last 2.5 days. Therefore, the amplitude of Fd and the first-step
decrease are mainly determined by sheath, while MC contributes to the formation of
the second-step decrease and the prolonged recovery time.

The magnetic turbulence and background magnetic field are the two distinguishing
features of sheath-MC structure, so their effects should be investigated. We carry 
out two simulations in which sheath-MC structure is included. In the first 
simulation the magnetic turbulence level is set according to Section~\ref{sec:para}
while background magnetic field is the same as that of ambient solar wind, with the
result presented by the green curve in Figure~\ref{fig:simulation_compare}(b). 
However, in the second simulation the background magnetic field is set according to
Section~\ref{sec:para} while magnetic turbulence level is the same as that of ambient
solar wind, with the result exhibited by the blue curve in 
Figure~\ref{fig:simulation_compare}(b). The other curves are the same as that of
Figure~\ref{fig:simulation_compare}(a). It is shown that the green curve has a
moderate gradual decrease after the shock arrival, with the recovery time about the
half of the red curve. Though the blue curve shows a two-step decrease, the amplitude
of the second step decrease is too small compared to the observation. The recovery
phase of the blue curve is also in line with that of the red curve in the last 2.5
days. It is shown that neither only magnetic turbulence nor background magnetic field
itself can produce the observed two-step decrease. Instead, the combination of
magnetic turbulence and background magnetic field in sheath-MC structure produced the
two-step Fd, and the enhanced background magnetic field in sheath-MC structure
extended the \edit1{total} recovery time.

\section{CONCLUSIONS AND DISCUSSION}
\label{sec:discussion}

In our previous work, WAQ-I, we use the sheath-MC model to numerically reproduce the
intensity-time profiles of relatively low energy SEPs in the GLE event on 2000 July
14 successfully. It is suggested that both SEP events and Fds are two important
phenomena accompanied with ICMEs, so that the results from compatible models should
be consistent with both SEP and GCR observations. In this paper, as the continuation,
we use the same sheath-MC model, same diffusion coefficients with magnetic turbulence
as input, and similar numerical method as in WAQ-I to reproduce Fds associated with
the same GLE. It is shown that there was a two-step decrease in the Oulu NM counting
rates when the sheath-MC structure arrived at the Earth. we assume the two-step Fd
was caused by the sheath and MC. Here, we solve the focused transport equation
instead of the Parker equation since focusing effect may be large in sheath-MC
structure. The simulation result is used to compare with the observed normalized
intensity of GCRs measured by the Oulu NM with the effective energy as 6.54 GeV.
Since Fd is a short term process and we only focus on the relative variation of GCRs,
the boundary for the GCR source is set to 10 au for reducing the consumption of
computing resources. Besides, the drift effect is neglected
\edit1{since it may have no significant influence on the formation of the two-step
decrease, amplitude, and total recovery time of the Fd}.

In the simulation model, the MC and sheath are set to spherical caps with fixed
thickness moving in a uniform speed. The Parker field is adopted as the background
magnetic field in solar wind, on which a magnetic enhancement in radial direction 
expressed by Equations~(\ref{eq:Bejecta}) is superposed as the background
magnetic field in sheath-MC structure. The magnetic turbulence levels in sheath and
MC are set to higher and lower than that in ambient solar wind, respectively.
The simulation result with sheath-MC structure can well reproduce the Fd event 
except for the first half of the recovery phase, and the two-step decrease occurs at 
the arrival of sheath and MC. The simulations with only one of sheath and MC infer
that sheath plays an important role in the amplitude of Fd while MC contributes to
the formation of the second step decrease and prolongs the recovery time of Fd. To 
evaluate the respective effects of magnetic turbulence and background magnetic field
in sheath-MC structure on Fd, we carry out simulations with only one of the magnetic 
turbulent level and background magnetic field set with the ejecta model with other
one set with ambient solar wind model in sheath-MC structure. The simulations show
that neither the magnetic turbulence nor the background magnetic field in sheath-MC
alone is sufficient to produce the observed two-step decrease in terms of the shape
and amplitude of Fd. Therefore, sheath and MC and their distinguishing features,
i.e., magnetic turbulence and background magnetic field are important for the
formation of the two-step Fd associated with GLE59.

In this work, the input parameters of the diffusion models, i.e., the turbulence
level, correlation length, and magnetic field power spectrum index are simplified.
In order to be consistent with WAQ-I, recent progress in turbulence theory
\citep[e.g.,][]{ZankEA18, ZhaoEA18, AdhikariEA20, ChenEA20}, especially the
radial dependence of turbulence parameters, is not included in this work. It is
supposed that the simplified turbulence parameters can be used to study the Fd, 
which is a local and short time scale phenomenon.

In general, the simulation result with sheath-MC structure captures the main features
of the observation, such as the pre-increase precursor, two-step decrease, amplitude,
\edit1{total} recovery time, etc. However, it is shown that the simulation deviates
from the observation in the first half of the recovery phase, which is reasonable due
to the simplifications we make in this work. Firstly, the shapes of sheath and MC are
set to thick spherical caps, which is different than the real ones. Secondly, the 
background magnetic field enhancement in sheath-MC structure is set in radial
direction, which is different from that of the real one. In addition, the background 
magnetic field enhancement model we use is not divergence free.
\edit1{Thirdly, the drift effect, which may influence the profile of recovery phase
of the Fd, is neglected.}
The further study can be carried out with the help of some new methods, e.g., the
Grad-Shafranov reconstruction technique \citep{HuASonnerup02, Hu17} or
magnetohydrodynamic \edit1{(MHD)} simulations
\citep[e.g.,][]{LuoEA13, PomoellEA18, WijsenEA19, Fend20} that can produce more
\edit1{realistic} sheath-MC structure.
\edit1{In addition, with the MHD simulation, which can produce a self-consistent HCS
with the existence of sheath-MC structure, we can investigate the drift effect.}
Finally, GCRs with high energy have large gyro-radius relative to the length scale
in which the background magnetic field varies significantly, so that GCRs may get
into sheath-MC structure through gyromotions, and the focus transport equation based
on the gyro-average of particles becomes invalid. To deal with such kind of problem
in the future, we may \edit1{analyze} large amount of trajectories of energetic
particles by numerically solving particles Newtonian equation of motion with Lorenz
force in interplanetary space \citep[e.g.,][]{QinEA02, KongEA17}.

\begin{figure}
\epsscale{1} \plotone{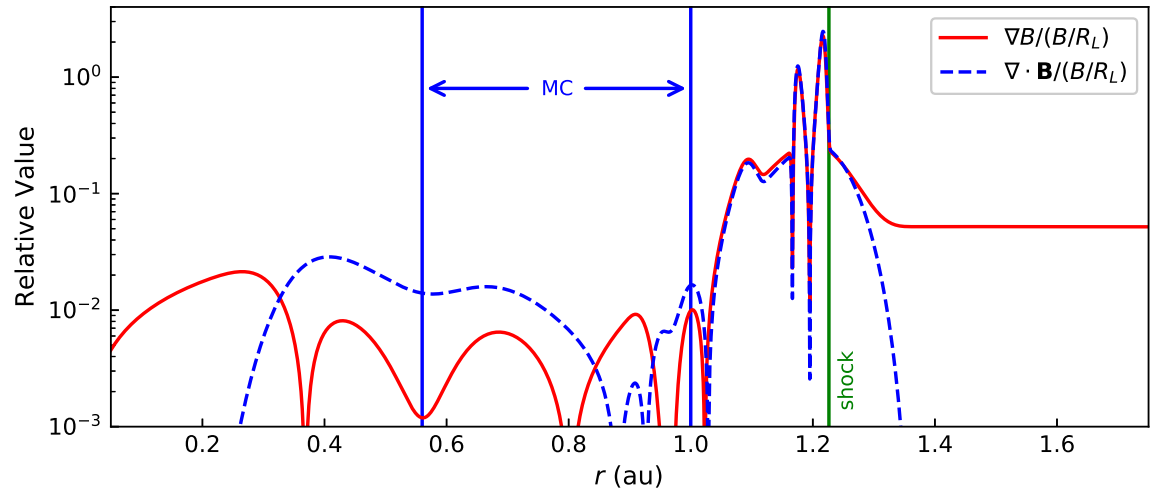}
\caption{The relative values of gradient and divergence of the modeled background
magnetic field in the Sun-Earth line versus the distance from the Sun at the time
of the modele MC arrival.}
\label{fig:divergence}
\end{figure}

\edit1{To evaluate the influence of the non-zero divergence, the relative divergence
of the modeled background magnetic field in the Sun-Earth line is presented in
Figure~\ref{fig:divergence} with the blue dashed line at the time of MC arrival.
In MHD simulations of solar wind, the relative divergence error of each cell in the
discrete space can be defined as $\frac{|\nabla \cdot \bm{B}|}{B/R_{\text{c}}}$
\citep[e.g.,][]{ZhangAFeng16}, where $R_{\text{c}}$ is the characteristic size of
the cell. The value of $R_c$ is about 0.01 au at 1 au, and the value is similar to
that of the Larmor radius $R_L$ of 6.54 GeV protons at 1 au. The relative divergence
error less than $10^{-2}$ is supposed small enough to simulate solar wind stably and
accurately \citep[e.g.,][]{ShenEA18}. In this work, we therefore define the relative
divergence as $\frac{|\nabla \cdot \bm{B}|}{B/R_{L}}$. Figure~\ref{fig:divergence}
shows that the relative divergence of the modeled background magnetic field is less
than or approximately equal to $10^{-2}$ except for the sheath region. In addition,
the relative gradient, defined as $\frac{|\nabla B|}{B/R_{L}}$, is also shown in
Figure~\ref{fig:divergence} with the red solid line. It is shown that the relative
divergence and gradient are of a similar order of magnitude. In the model of GCRs, a
uniform local background magnetic field is assumed, so that the non-zero gradient
introduces error in the GCR modeling. We suppose the error of the GCR modeling from
non-zero divergence of magnetic field has the similar level as that from the non-zero
gradient of magnetic field in sheath region.}

\acknowledgments
This work was supported, in part, under grants NNSFC 41874206 and NNSFC 42074206.
We thank the \textit{Wind} MFI teams for providing the data used in this paper.
We appreciate the availability of the \textit{Wind} data at the Coordinated Data
Analysis Web. The cosmic ray data were supplied by the Neutron Monitor Database
(NMDB), which was founded under the European Union's FP7 program
(contract No. 213007). The work was carried out at National Supercomputer Center
in Tianjin, and the calculations were performed on TianHe-1 (A).




\clearpage

\clearpage

\clearpage

\clearpage

\clearpage

\clearpage
\begin{deluxetable}{cccc}
\tablecaption{Parameter settings for the simulation.
\label{tab:para}}
\tablehead{
\colhead{Type} & \colhead{Parameter} & \colhead{Meaning} & \colhead{Value}
}
\startdata
\multirow{2}{*}{Shock} & $v_{\text{s}}$ & speed & 1406 km/s\\
& $\Omega_{\text{s}}$ & half angular width & $45^{\circ}$\\
\hline
\multirow{3}{*}{Solar wind} & $V^{\text{sw}}$ & speed & 450 km/s\\
& $B_{\text{P}0}$ & radial strength of IMF at 1 au & 3.62 nT\\
& $B_{\text{P}}|_{1\text{au}}$ & total strength of IMF at 1 au & 5 nT\\
\hline
\multirow{2}{*}{MC} & $L_{\text{M}}$ & half thickness & 0.22 au\\
& $d_{\text{M}}$ & distance between MC center and shock & 0.45 au\\
\hline
\multirow{1}{*}{Sheath} & $L_{\text{S}}$ & half thickness & 0.08 au\\
\hline
\multirow{4}{*}{Others} & $\omega$ & angular speed of solar rotation & 
  2$\pi$/25.4 rad/day\\
& $R_{\text{in}}$ & inner boundary & 0.05 au\\
& $R_{\text{out}}$ & outer boundary & 10 au\\
& $E_{\text{eff}}$ & effective energy of Oulu NM & 6.54 GeV\\
\enddata
\end{deluxetable}

\clearpage
\begin{deluxetable}{cccc}
	\tablecaption{Parameter settings of turbulence for the simulation.
		\label{tab:para_turbulence}}
	\tablehead{
		\multicolumn{2}{c}{Parameter} & \colhead{Meaning} & \colhead{Value}
	}
	\startdata
	\multicolumn{2}{c}{$\sigma_{\text{P}}$} & turbulence level in solar wind & 0.3\\
	\hline
	\multirow{2}{*}{$\sigma_{\text{ejecta}}$} & $\sigma_{\text{S}}$ & turbulence level in sheath & 1.6\\
	& $\sigma_{\text{M}}$ & turbulence level in MC & 0.1\\
	\hline
	\multirow{2}{*}{$l_{\text{turb}}$} & {$l_{\text{slab}}$} & slab correlation length & 0.025 au\\
	& {$l_{\text{2D}}$} & 2D correlation length & 0.0096 au\\
	\hline
	\multicolumn{2}{c}{$s$} & Kolmogorov spectral index &$5/3$\\
	\hline
	\multicolumn{2}{c}{$h$} &  non-linear effect index & $0.01$\\
	\enddata
\end{deluxetable}

\end{document}